# INTELLIGENT STRATEGIES FOR DAG SCHEDULING OPTIMIZATION IN GRID ENVIRONMENTS

**Florin Pop, Valentin Cristea**

*Faculty of Automatics and Computer Science, University "Politehnica" of Bucharest*
*{florinpop, valentin}@cs.pub.ro*

Abstract: The paper presents a solution to the dynamic DAG scheduling problem in Grid environments. It presents a distributed, scalable, efficient and fault-tolerant algorithm for optimizing tasks assignment. The scheduler algorithm for tasks with dependencies uses a heuristic model to optimize the total cost of tasks execution. Also, a method based on genetic algorithms is proposed to optimize the procedure of resources assignment. The experiments used the Mon*ALISA* monitoring environment and its extensions. The results demonstrate very good behavior in comparison with other scheduling approaches for this kind of DAG scheduling algorithms.

## 1. INTRODUCTION

Grid is the subject of many international research projects in Europe (for example EGEE, SEE-GRID, *etc*.) and in US (e.g. OSG). One of the main actual requirements is finding solutions for efficient and high performance execution of applications that are computing intensive, data intensive or a combination of both. The answers include breaking the problem into smaller pieces, which means partitioning the jobs into smaller tasks, discovery of available services and resources, scheduling tasks and workflows and distributing them to specific system nodes, providing the data where and when they are required, collecting results and giving them to the interested user. In addition, execution management assists the decision processes in the Grid environments, which are based on autonomic features such as self-configuration, self-optimization, self-recovery, and self-management (Joshy *et al*, 2004).

The number of clusters in the global Grid, the heterogeneity of resources and the complexity of the applications require an efficient Grid scheduling solution. The Grid scheduler should cope with receiving job execution requests from users, receiving information about the available resources / services, and mapping the application to resources and services according to some optimization criteria. Several problems are related with this approach.

First, a Grid scheduler cannot control the clusters/resources directly. It is more natural to consider that Grid schedulers are closely related to Grid applications, and are responsible for the management of jobs, such as allocating resources needed for any specific job, managing the tasks for parallel execution, managing of data, and correlation of events. One solution is to view Grid scheduling as a Web Service, which uses other Web Services (for example a Grid Information Service) to fulfill its function. This solution has several advantages. First, it can support a more complex utilization of Grids, such as the coordinated resource sharing, and execution cost optimizations for applications. Then, Grid applications can achieve levels of flexibility utilizing infrastructures provided by middleware frameworks that use the web services concept. Globus Toolkit 4 is the most known implementation of a flexible web services framework (Foster, 2005) for Grid applications. Using this framework it is possible to create scheduling services for all type of applications and for large Virtual Organizations as well. The LHC physics project (for nuclear physics experiments) provides the main applications for Grid (LCG, 2007). LCG (LHC Computing Grid) which is based on Globus are the main middleware used in LHC.

Second, the partition of a job into tasks can lead to dependent tasks. Obviously, this asks for the design and use of more complex scheduling algorithms.

Third, more than one single criterion could be required for scheduling optimization. New approaches could be needed to accommodate such claims.

In this paper we present a method for optimizing the scheduling mechanism for dependent tasks. The paper is structured as follows: Section 2 is a general presentation of the task DAG Scheduling methods and open issues. Section 3 describes the structure and functionality of the proposed solution. Section 4 introduces the main implementation issues. We describe and comment on the results in the 5th section. Section 6 contains conclusions and directions for future research.

## 2. GENERAL PRESENTATION OF THE DAG OF TASKS SCHEDULING

As mentioned above, partitioning a job into tasks can lead to dependent tasks. A dependent task cannot start before the execution of the tasks it depends on is terminated. To represent a set of task and their dependencies we can use a directed acyclic graph (DAG) where the tasks are represented by nodes and the dependencies are represented by arcs. If starting the execution of the task B depends on terminating the execution of the task A then an arc (directed edge) from A to B exists in the associated DAG. Obviously, one task may depend on several other tasks.

Tasks' dependencies have a major role in the design of scheduling algorithms. The taxonomy of Grid scheduling algorithms for dependent tasks is shown in Figure 1 (Dong *et al*, 2006). In the sequel, by *DAG scheduling* we denote the scheduling of tasks with dependencies.

### 2.1. Properties of the DAG scheduling problem

The general DAG scheduling problem is NP-complete (Yu-Kwok *et al* 1997, 1999, Kohler *et al*, 1976). An approximation algorithm could aim to provide a polynomial time solution for some particular cases.

One case is that of a tree-structured graph with identical computation costs for the tasks. A linear time solution exists for scheduling these tasks on an arbitrary number of processor. (Hu, 1961).

Another case is to schedule arbitrary graphs with identical task computation costs, on two processors.

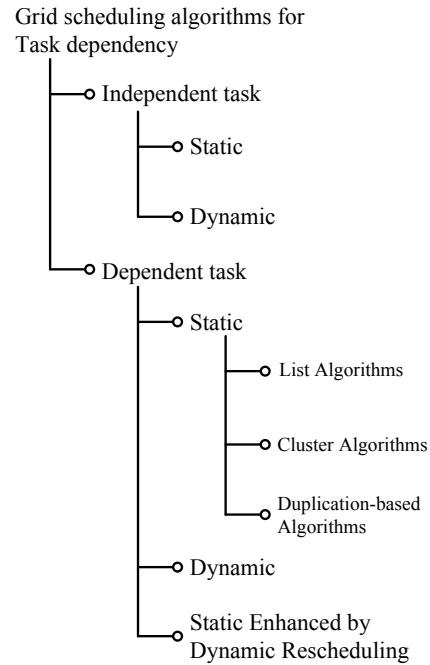

Figure 1. Taxonomy of task dependency scheduling algorithms in Grid environments

The solution is a quadratic-time algorithm (Coffman *et al*, 1972).

The third case we mention is the scheduling of the interval-ordered DAG with uniform node weights to an arbitrary number of processors. Again, a linear time solution exists for this case (Papadimitroiu *et al*, 1979).

### 2.2. Background of DAG scheduling problem

A task graph is a directed acyclic graph $G(V, E, c, \tau)$ where:
- $V$ is a set of nodes (tasks);
- $E$ is a set of directed edge (dependencies);
- $c: V \to R_+$ is a function that associates a weight $c(u)$ to each node $u \in V$; $c(u)$ represent the execution time of the task $T_u$, which is represented by the node $u$ in $V$;
- $\tau$ is a function $\tau: E \to R_+$ that associates a weight to a directed edge; if $u$ and $v$ are two nodes in $V$ then $\tau(u, v)$ denotes the inter-tasks communication time between $T_u$ and $T_v$.

A task graph example is shown in Figure 2. Two special kinds of nodes can be identified. A *source node* is a node without incoming directed edges. It corresponds to the task that initiate the entire application, and it is the first task in any possible schedule. An *exit node* is a node without outgoing directed edges. Two items are associated with each node: the *task id* $T_u$ is represented in the upper half of the node (circle), while the *execution time* $c(u)$ is represented in the lower half. Each edge is labeled with the inter-tasks communication time.

If we denote $st(u)$ the *start time* and $ft(u)$ the *finish time* for task $u$, and define

$$makespan = \max_{u \in V}\{ft(u)\},$$

we can formulate the goal of optimizing DAG scheduling as follows: minimize the *makespan* without violating precedence constrains.

A scheduler is considered *efficient* if the *makespan* is short and respects resource constrains, such as a limited number of processors, memory capacity, available disk space, *etc*.

Many types of scheduling algorithms for DAG are based on the list scheduling technique. Each task has an assigned priority, and scheduling is done according to a list priority policy: select the node with the highest priority and assign it to a suitable machine. According to this policy, two attributes are used for assigning priorities:
- *tlevel* (top-level) for a node $u$ is the weight of the longest path from the source node to $u$.
- *blevel* (bottom-level) for a node $u$ is the weight of the longest path from $u$ to an exit node.

The time-complexity for computing *tlevel* and *blevel* is $O(|V|+|E|)$.

We define the *ALAP* (*A*s *L*ate *A*s *P*ossible) attribute for a node $u$ to measure how far the node's start-time, $st(u)$ can be delayed without increasing the *makespan*. This attribute will have an important role for load balancing constrains because it show if we can delay the execution start of a task $T_u$.

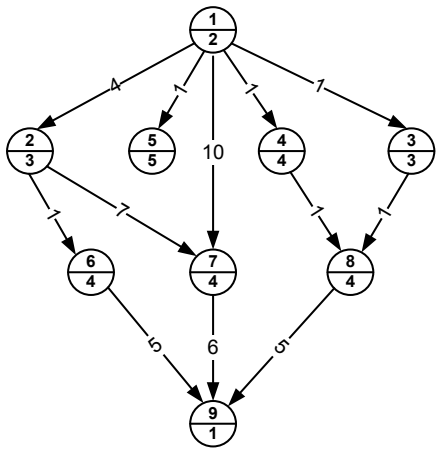

Figure 2. Task DAG Example

For the example in Figure 2, the *tlevel*, *blevel* and *ALAP* parameters are presented in Table 1. For the node 1 (task $T_1$) the *tlevel* is 0. It means that this node must be scheduled and executed without any time delay. The *blevel* is 23 and denotes the *makspan* for this graph. For the node 9 (task $T_9$) the *ALAP* is 22 (the maximum time before task 9 ca start the execution).

Table 1. The *tlevel, blevel* and *ALAP* for Figure 2 graph example

| Node | tlevel | blevel | ALAP |
|---|---|---|---|
| 1 | 0 | 23 | 0 |
| 2 | 6 | 15 | 8 |
| 3 | 3 | 14 | 9 |
| 4 | 3 | 15 | 8 |
| 5 | 3 | 5 | 18 |
| 6 | 10 | 10 | 13 |
| 7 | 12 | 11 | 12 |
| 8 | 8 | 10 | 13 |
| 9 | 22 | 1 | 22 |

### 3. HEURISTIC MODEL BASED ON GENETIC ALGORITHMS FOR GRID SYSTEMS

Grid scheduling for DAG task dependencies involves the main phases specified in (Schopf, 2003) to which we added the new *DAG tasks scheduling* phase that prepares the tasks for submission.

- *Resources discovery*. This phase generates a list of potential resources with corresponding specifications (available number of processors, free memory or free disk *etc*.). In this phase it is possible to make resource reservations (this is one of the particularities used in grid economy model). For task scheduling the resource characteristics are important because they permit the allocation of several tasks on the same resource (e.g. one cluster) if the resource characteristics allow this. These recourse parameters are provided by the Grid Information System (GIS).

- *DAG tasks scheduling*. In this phase, the tasks are prepared for submission to different systems in Grid.

- *System selection*. In this phase the best set of resources are selected from the sets of resources obtained in the first phase.

- *Job execution*. The tasks are mapped to the selected resources and are submitted for execution.

The algorithms based on *makespan* minimization heuristic are the ones giving the best results. The main purpose is to provide a lower bound for *makespan*. To this respect, each task is assigned the earliest possible time to start the execution. This is the best solution for the *List algorithms* class (see Figure 1).

But, during the execution, the *makespan* can change due to modifications in the order of task executions and/or to changes in inter-task communication times. This scenario describes a dynamic algorithm.
A dynamic algorithm is CCF (Cluster ready Children First). In this algorithm the graph is visited in topological order, and tasks are submitted as soon as scheduling decisions are taken (Forti A., 2006).

In the CCF algorithm, when a task is submitted for execution it is inserted into RUNNING-QUEUE (see Figure 3). If a task is extracted from the RUNNING-QUEUE, all its successors are inserted into the CHILDREN-QUEUE. These queues can be priority queues where the priority is assigned based on *tlevel* and *blevel* parameters. For example the priority for each task can depend on the $tlevel(u) + blevel(u)$ factor.

The outline of the CCF algorithm is presented bellow (in this algorithm *deq* is dequeue operation and *enq* is the operation enqueue).

---

computes *tlevel* and *blevel* for each node;

insert source task into RUNNING-QUEUE;

while (! isEmpty (RUNNING-QUEUE))

   *task* = deq (RUNNING-QUEUE);

   for-each *child* of *task* do
        enq (*child*, CHILDREN-QUEUE);
   end for

   while (! isEmpty (CHILDREN-QUEUE))

      *child_task* = deq (CHILDREN-QUEUE);

      if (isReady (*child_task*)) then

         assignResource (*child_task*);
         updateResources (*child_task*);
         enq (*child_task*, RUNNING-QUEUE);

      else
         suggestedResources (child_task);
      end if

   end while
end while

---

In Figure 3, the Monitoring System is the goal of the *updateResource* function. This goal can be achieved through using a monitoring system (e.g. Mon*ALISA*, Ganglia *etc*.), and a good scheduling strategy applied to the local level (clusters) and global level (entire system) of Grids. (Pop F. et al, 2006). The monitoring system can offer information on tasks execution in the Grid environment (Grid Clusters) as a feed-back to scheduler.

The central part of the algorithm is the *assignResource* function. This function considers two important aspects:
- the set of candidate resources composed by the resource assigned by the initial static mapping of tasks, the resources of the task parents, and one suggested resource;
- the scheduling target, which is to minimize the cost function.

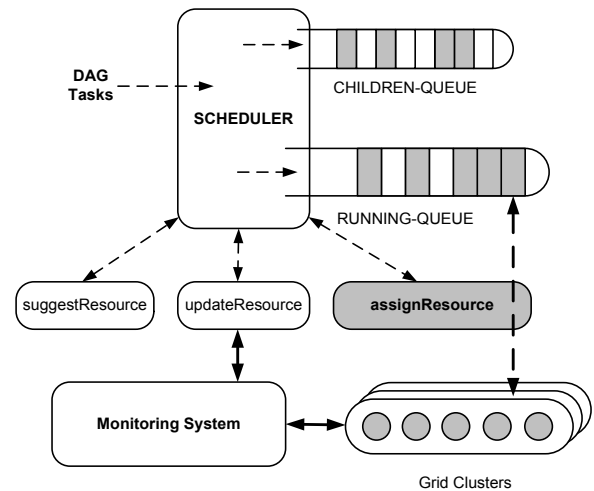

Figure 3. Scheduler Architecture

The resources assigned can be determined for the RUNNING-QUEUE (this queue contains tasks without dependencies) under the constraints of cost that have to be minimized.

### 4. OPTIMIZATION METHOD FOR DAG TASKS SCHEDULING ALGORITHM

We propose a method based on genetics algorithms for resources assignment. This method (Iordache et al, 2006) uses genetic algorithms (GAs) to compute a fitness function for each task and schedule the set of tasks using these values. We proposed and implemented this method for the *assignResource* function in the CCF algorithm.

The description of the scheduling method is presented in the following, in a logical flow of activities:

- A user requests that one task is scheduled. This request has as a parameter the name of the file

- containing a description of the tasks. The file has a standard XML format and presents task requirements related to memory, processor usage, execution time, etc.
- The input file is processed and a "batch of tasks" (group of tasks) objects is constructed for independent task from RUNNING-QUEUE.
- The batch of tasks is broadcast to all the nodes in the cluster.
- The nodes receive the group of tasks to be scheduled. The tasks are inserted in a queue by its priority according to a fitness function. If the number of tasks in the queue is less than a predefined length of the chromosome (in the solution, a fixed length of chromosome is used), they wait for $T$ units of time before starting the genetic algorithm. If the chromosome is still not complete at the end of the waiting period, a non-influential padding is added. On the contrary, if the length of an arriving group of tasks exceeds the predefined dimension of the chromosome, some tasks are saved in the waiting queue and will be scheduled at a latter time.
- On each node, up-to-date information on the status of the computers in the Grid on which tasks is sent for execution is kept, by constantly queering a monitoring system. The nodes query the Monitoring System, at the beginning of the algorithm, to find out the current status of the Grid.
- Each node starts with a different, specific initialization of the GA. The subsequent steps of the GA are similar for all the nodes in the cluster, and the same fitness formula is used. In this way, the clients will compute different optima from which the best one will be chosen.
- The migration of the best current solutions is performed after each step of the GA, thus ensuring that the population finds a optimal solution. The nodes exchange the best individuals and include them in the next generation.
- The generation of populations ends after a finite, predefined number of steps. At this point, each client in the cluster computes its optimal solution.
- The same communication procedure as above is used for the final step of the GA. Each node sends its optimum to all the other nodes in the cluster and the final optimal individual is decided by each of them. The best chromosome is selected from the optimal individuals. The result is the same on every node, because the computing procedure and the individuals at the last step of the GA are the same.
- The scheduling obtained is saved in a history file on each node in the cluster.

This is a distributed, fault-tolerant, scalable, and efficient solution of dynamic scheduling for optimized assignment of tasks from RUNNING-QUEUE or CHILDREN-QUEUE. The *assignResource* uses a combination of genetic algorithms and monitoring services for obtaining a scalable and highly reliable optimization tool. The monitoring system used is the Mon*ALISA* environment and its extensions. As we'll see in the result section, the results highlight very good behaviour in comparison with other decentralized scheduling approaches.

## 5. EXPERIMENTAL RESULTS

The testing of CCF algorithm on the example in Figure 2 led to the results presented in Figure 4.

We compute the cost (*c*) of the task according to its requirements: needed memory, CPU speed (given in MIPS), *etc*. To obtain the inter-task communication time ($\tau$) we have to retrieve the link characteristics (latency and bandwidth) using the formula:

$$\tau = \frac{data\_size}{bandwidth} + latency \ .$$

The communication computation ratio is defined as the average edge weight divided by the average node weight.

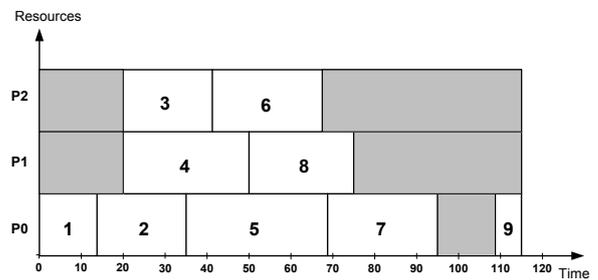

Figure 4. Result of CCF algorithm

The experiment for the *assignResource* analyses the metrics discussed in a decentralized scheduling scenario (Iordache G. et al, 2006), in which a cooperative genetic strategy has been employed. The cooperative characteristic implies optimal individuals interchange in order to speed up convergence. The input task set is the same as previously, as well as the level of decentralization (3), and the number of generations (100).

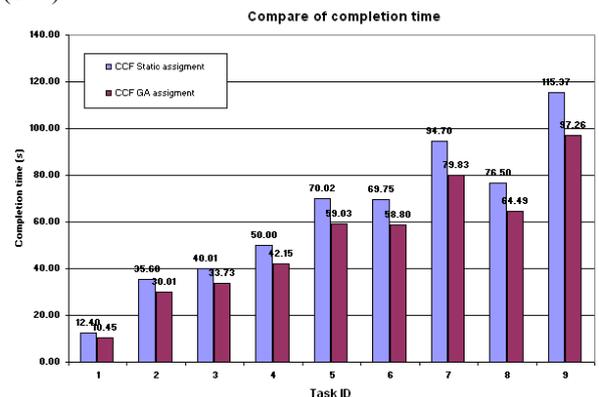

Figure 5. Comparison the completion time between two strategies for resource assignment

Figure 5 illustrates the schedule obtained. An essential improvement of completion time has been achieved, namely an improvement of 16%, in comparison with the static strategy for assigning resources. This performance comparison of the proposed method and static assignment methods is made using tree processors (as we see in Figure 4). Tests were performed using a graph with nine tasks (example from figure 2).

For the entire set of tasks (with dependencies) we obtained a good load balancing for processors if the communication time is low and the task dependencies have no many edges. A good load balancing is obtained if we consider only the task from RUNNING-QUEUE (with no dependencies).

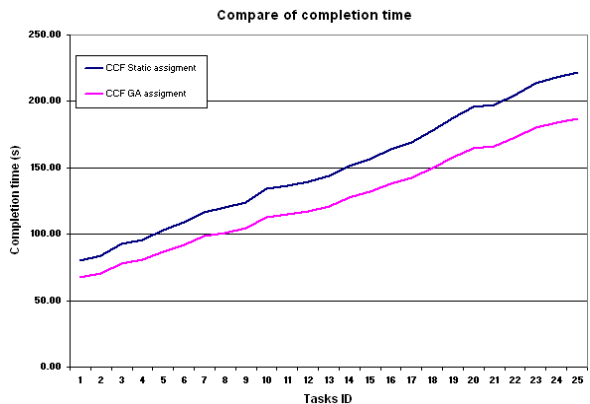

Figure 6. Completion time compare for 25 dependent tasks

Finally, we tested the resource assignment for 25 dependent tasks. The completion time is shown in Figure 6. An improvement of 16% for completion time has been achieved.

## 6. CONCLUSIONS AND FUTURE WORK

Task dependencies are more frequently encountered in Grid applications. Scheduling solutions for these cases are required under the constraints of QoS. We proposed a heuristic solution for existing DAG scheduling algorithm, namely: cluster ready, children first. The solution uses genetic algorithms for the resource assignment process. With this approach, we obtained a good improvement in QoS as compared with the static assignment method.

The main contribution of this research is the use of an intelligent, heuristic method for optimizing Grid scheduling for DAG. Future work will consider: new scheduling algorithms for real-time scenarios, solutions for backup and recovery from error (re-scheduling), optimized file transfer, solving the problem of co-scheduling, multi-criteria constrains scheduling.
We also want to include this intelligent method for DAG scheduling in the DIOGENES (DIstributed Optimal GENEtic Algorithm for grid application Scheduling) project (DIO, 2007).

## REFERENCES

Armstrong, R., Hensgen, D., Kidd, T. *The relative performance of various mapping algorithms is independent of sizable variances in run-time predictions*. Procs. of the 7th IEEE HCW, pp. 79-87, 1998.

Cao, J. et al. *Grid load balancing using intelligent agents.* Future Generation Computer Systems special issue on Intelligent Grid Environments: Principles and Applications, 2004.

Coffman, E.G., Grahman, R.L., *Optimal scheduling for two-processor systems.* Acta Informatica, 1:200-213, 1972.

DIOGENES project (DIO), http://diogenes.grid.pub.ro/, Accessed on 15 January 2007.

Dong, F., Akl, S.G., *Scheduling Algorithms for Grid Computing: State of the art and opened problems*, Technical report no. 2006-504, January 2006.

EGEE project, http://public.eu-egee.org, Accessed on 15 January 2007.

FORTI A., *DAG Scheduling for Grid Computing systems*. Ph.D. Thesis, University of Udine – Italy, 2006

Foster, I., *Globus Toolkit Version 4: Software for Service-Oriented Systems*, in the Proceedings of IFIP International Conference of Network and Parallel Computing, Springer-Verlag LNCS Vol. 3779, pp 2-13, Beijing, China. December 2005.

Heymann, E., Fernndez, A., Senar, M.A., Salt J. *The EU-CrossGrid Approach for Grid Application Scheduling*. LNCS, Vol. 2970, pp. 17-24, 2004.

Hu, T.C., *Parallel sequencing and assembly line problems. Oper. Research.* 19(6):841-848, November, 1961

Iordache G., Boboila M., Pop F., Stratan C., Cristea V., *A Decentralized Strategy for Genetic Scheduling in Heterogeneous Environments*, Proceedings of GADA Conference, LNCS 4276 proceedings, pp. 1234-1251, Montpellier, France, November 2-3, 2006.

Jenifer M. Schopf, *Ten Action when scheduling*, In Grid resource management: state of the art and future trends, chapter 2, pages 15-23, Kluwer Academic Publishers, 2003.

Joshy, J., Fellenstein, C., *Introduction to Grid Computing,* Prentice Hall PTR, Apr 16, 2004.

LCG project, http://lcg.web.cern.ch/LCG, Accessed on 15 January 2007.

Papadimitroiu, C.H., Yannakakis, M., *Scheduling interval-order tasks.* SIAM J. Computing, 8:405-09, 1979.

Pop F., Dobre C., Godza G., Cristea V., *A simulation model for grid scheduling analysis and optimization*, Proceedings of PARELEC Conference, pp. 133-138, Bialystok, Poland, September 13-17, 2006, ISBN: 0-7695-2554-7.

SEE-GRID project http://www.see-grid.org, Accessed on 15 January 2007.

Yu-Kwong Kwok, *High-performance algorithms of compile-time scheduling of parallel processors.* PhD thesis, Hong Kong University of Science and Technology, 1997

Yu-Kwong Kwok and Ishfaq Ahmed, *Static scheduling algorithm for allocating directed task graph to multiprocessors.* ACM Computing Surveys 31(4):406-471, 1999.